\documentclass[journal]{IEEEtran}
   
\usepackage{bbm}

\usepackage{tabularx}
\usepackage{amsthm}
\usepackage{epstopdf}
\usepackage{cite}
\usepackage{bbm}
\usepackage{amsmath,amssymb,amsfonts}
\usepackage{algorithm} 
\usepackage{algpseudocode} 
\usepackage{graphicx}
\usepackage{textcomp}
\usepackage{xcolor}
\usepackage[utf8]{inputenc} 
\usepackage[T1]{fontenc}    
\usepackage{hyperref}       
\usepackage{url}            
\usepackage{booktabs}       
\usepackage{amsfonts}       
\usepackage{nicefrac}       
\usepackage{microtype}      
\usepackage{wrapfig}
\usepackage{color}
\usepackage{float}
\usepackage{listings}
\usepackage{amsmath}
\usepackage{graphicx}
\usepackage{makecell}
\usepackage{tabularx}
\usepackage{caption}%
\usepackage{nopageno}
\usepackage{subcaption}
\usepackage{amsmath}
\usepackage{algorithm}
\usepackage{algpseudocode}
\usepackage{amssymb}
\usepackage{enumitem}
\makeatletter
\def\ps@headings{%
\def\@oddhead{\mbox{}\scriptsize\rightmark \hfil \thepage}%
\def\@evenhead{\scriptsize\thepage \hfil \leftmark\mbox{}}%
\def\@oddfoot{}%
\def\@evenfoot{}}
\makeatother
\usepackage{caption}
\usepackage[letterpaper, left=0.641in, right=0.641in, bottom=1.095in, top=0.73in]{geometry}
\newcolumntype{A}{>{\centering\arraybackslash}p{0.3\textwidth}}
\newcolumntype{B}{>{\centering\arraybackslash}p{0.17\textwidth}}
\usepackage{framed}

\begin{document}

\title{
Delay-Aware Reinforcement Learning for O-RAN Control under Stochastic Reward Feedback
}

\author{Xingqi Wu, \emph{Student Member IEEE}, Junaid Farooq, \emph{Senior Member IEEE}, Tao Li, \emph{Member IEEE}, \\ Juntao Chen, \emph{Senior Member IEEE}, and Ying Wang, \emph{Member IEEE} \vspace{-0.2in}
\thanks{X. Wu and J. Farooq are with the Department of Electrical and Computer Engineering, University of Michigan-Dearborn, MI 48128 USA, Emails: \{xingqiwu, mjfarooq\}@umich.edu.}
\thanks{J. Chen is with the Department of Computer and Information Science, Fordham University, New York, NY USA, Email: jchen504@fordham.edu.}
\thanks{T. Li is with the Department of Systems Engineering, City University of Hong Kong, Hong Kong., Email: li.tao@cityu.edu.hk.}
\thanks{Y. Wang is with the Department of Systems Engineering, Stevens Institute of Technology, Hoboken, NJ USA, Email: ywang6@stevens.edu. }
}

\maketitle
 
\begin{abstract}
Reinforcement learning (RL) is increasingly employed for real-time control in Open Radio Access Networks (O-RANs). However, the asynchronous nature of O-RAN feedback introduces stochastic delays between actions and their observable rewards, leading to misaligned credit assignment and unstable training. This letter formulates the O-RAN control problem as a Markov decision process (MDP) with delayed and stochastic rewards, where the true reward corresponding to an action is received after an unknown delay. To address this, we propose a delay-aware deep RL framework that integrates a long short-term memory (LSTM)-based delay predictor and a temporal reward realignment module within an off-policy learning pipeline. Using empirical delay traces collected from the Open AI Cellular (OAIC) testbed, we demonstrate that the proposed framework achieves faster convergence and significantly reduces latency violations and packet drops compared to conventional RL methods that ignore feedback delay. The results confirm that reward-delay compensation is critical for reliable AI-driven O-RAN control.
\vspace{-0.1in}
\end{abstract}


\IEEEpeerreviewmaketitle
\vspace{-0.0in}
\section{Introduction}

The open radio access network (O-RAN) paradigm promotes the disaggregation and virtualization of radio access functions, enabling the deployment of artificial intelligence (AI)-driven control through the near-real-time RAN Intelligent Controller (RIC)~\cite{RIC,ran-rl}. Reinforcement learning (RL) has emerged as a promising tool to optimize such control processes, allowing agents to adaptively manage radio resources, slicing, and admission policies in highly dynamic environments~\cite{AC-NS_optimization-RL-5G,MADRL_o_ran,AC-Dynamic-MPC}.

However, while the use of RL for O-RAN control has been widely studied in simulation, practical deployments face a critical yet often overlooked challenge, i.e., stochastic delay in feedback signals. When an RL agent issues a control action (e.g., modifying physical resource block allocations or scheduling priorities), its effect on observable performance metrics such as throughput or latency does not manifest immediately due to queuing, MAC scheduling, and telemetry reporting cycles~\cite{oaic,O-ran-latency-interface}. From the perspective of learning, these effects appear as delayed rewards, where the agent receives feedback corresponding to a past action after an uncertain interval. Such delays distort the temporal association between actions and outcomes, corrupt the replay buffer, and degrade policy convergence.

 \begin{figure}[t]
    \centering
    \includegraphics[trim=0 0 0 0, clip, width=3.2in]{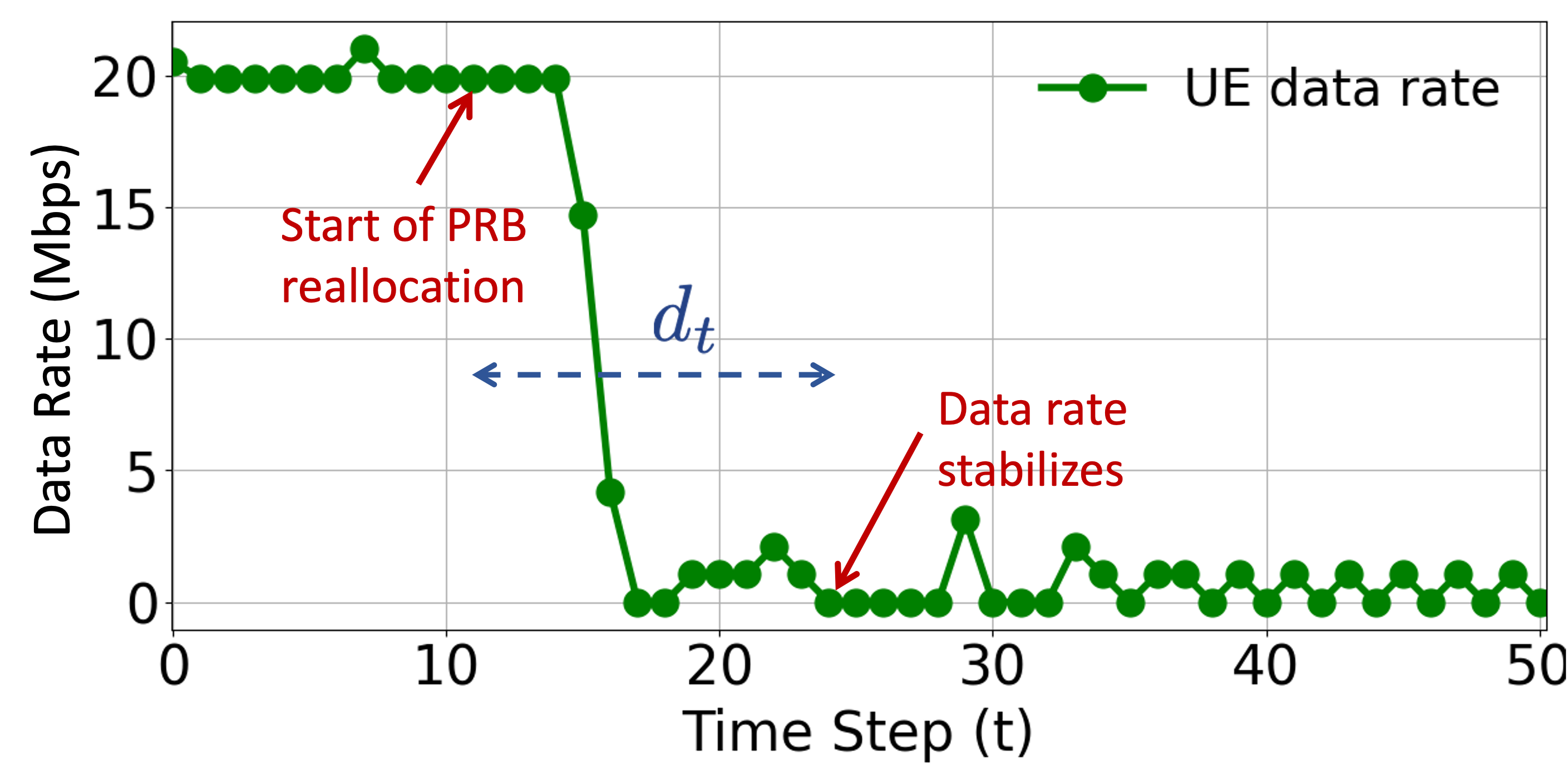}
    \caption{ \small Empirical observation of delayed feedback in the OAIC O-RAN testbed. The PRB allocation is reduced to zero at time $t$, but the downlink data rate only drops after a delay interval $d_t$. }
        \label{fig:trace}
\end{figure}
Empirical evidence of this phenomenon is shown in Fig.~\ref{fig:trace}, based on measurements from the Open AI Cellular (OAIC) O-RAN testbed~\cite{oaic}. When the PRB allocation is reduced to zero, the downlink data rate does not decrease instantaneously but only after some delay, highlighting the stochastic feedback latency inherent in real O-RAN control loops.
Prior works on RL-driven O-RAN control have studied problems such as network slice admission~\cite{AC-NS_optimization-RL-5G}, admission among heterogeneous slices~\cite{AC-max-Long-Term-Average-Revenue}, throughput-maximizing admission in edge computing~\cite{AC-DRL-throughput-max}, and multi-agent O-RAN resource allocation~\cite{MADRL_o_ran}. However, most of these approaches rely on offline training in simulators where feedback is synchronous. This assumption does not hold in real deployments, where delays are inevitable. On the other hand, the RL theory community has investigated delayed rewards in online learning and bandit settings~\cite{joulani2013online,quanrud2015advdelay}, off-policy learning with delayed feedback~\cite{hc-off-policy}, and reward decomposition methods~\cite{aggarwal_delay_rl,Rudder}. These studies demonstrate fundamental impacts of delayed feedback, but generally assume the delay is fixed or provided to the learner. Such assumptions are impractical in O-RAN, where delays vary unpredictably across implementations and must be inferred from observed system behavior.

To address this limitation, this letter introduces a delay-aware deep reinforcement learning (DRL) framework for O-RAN, designed to learn effectively under stochastic and unpredictable reward delays. The proposed framework integrates an LSTM-based delay predictor with an off-policy RL agent to realign rewards with their originating state–action pairs before policy updates. This enables the agent to recover the correct temporal credit assignment despite delayed feedback.
The key contributions of this work are as follows:
\begin{itemize}[topsep=1pt, partopsep=0pt, itemsep=1pt, parsep=0pt]
\item We formulate O-RAN control as a Markov decision process (MDP) with stochastic delayed rewards, capturing the effect of non-Markovian feedback caused by network control latency.
\item We propose a delay-aware deep RL framework that uses an LSTM-based predictor to estimate reward delay and perform temporal realignment during off-policy training.
\item We validate the framework using delay traces derived from the OAIC testbed, demonstrating superior stability and performance over conventional RL baselines.
\end{itemize}

\section{System Model and Problem Formulation}

We consider an O-RAN downlink system where a near-RT RIC employs a RL agent to optimize radio resource management across heterogeneous user slices. The network consists of a centralized unit (O-CU), a distributed unit (O-DU), and a radio unit (O-RU) interconnected via the F1 and E2 interfaces. The near-RT RIC communicates control actions to the O-DU through the E2 interface. The O-DU executes physical-layer and MAC-layer scheduling decisions, which in turn affect the transmission performance experienced by the user equipments (UEs).

Let $\mathcal{M} = \{1,2,\dots,M\}$ denote the set of active UEs, each belonging to either the enhanced Mobile Broadband (eMBB) or ultra-reliable low-latency communication (URLLC) slice, represented by $\mathcal{M}_{\text{e}}$ and $\mathcal{M}_{\text{u}}$, respectively. Each slice type enforces a distinct maximum tolerable latency $\tau_{\text{e}}^{\max}$ or $\tau_{\text{u}}^{\max}$ according to its quality-of-service (QoS) class.

At each decision epoch $t\in \mathcal{T}=\{0, \ldots, T_{n-1}\}$, the near-RT RIC observes a network state vector
$s_t = \{ \bar{\tau}_{\text{e}}(t), \bar{\tau}_{\text{u}}(t), \bar{b}_{\text{e}}(t), \bar{b}_{\text{u}}(t)\}$,
where $\bar{\tau}_{x}(t)$ and $\bar{b}_{x}(t)$ represent the average latency and throughput for slice $x \in \{\text{e}, \text{u}\}$. Based on $s_t$, the agent issues a control action
$a_t = \{ B_m(t), C_m(t) \}_{m \in \mathcal{M}}$,
where $B_m(t)$ and $C_m(t)$ represent the bandwidth and computational resource allocated to UE $m$, subject to $\sum_{m \in \mathcal{M}} B_m(t) \leq B$ and $\sum_{m \in \mathcal{M}} C_m(t) \leq C$.

\subsection{Latency and Queueing Model}

Each UE generates data packets that must be transmitted and processed before their latency deadline. The end-to-end latency $\tau_m(t)$ is composed of transmission and computation delays:
\begin{equation}
\tau_m(t) = \tau_m^{\text{tx}}(t) + \tau_m^{\text{comp}}(t).
\end{equation}
Both components are determined by the service rates induced by allocated bandwidth and compute capacity.
Let $\delta_m(t)$ and $N_m(t)$ denote the average packet size and arrival rate (packets per slot) for UE $m$. The instantaneous achievable data rate, given by the Shannon model, is
\begin{equation}
b_m(t) = B_m(t) \log_2\!\left(1 + \frac{P |h_m(t)|^2 \beta_m(t)}{B_m(t) N_0 + I_m(t)}\right),
\end{equation}
where $P$ is the transmit power, $|h_m(t)|^2$ the small-scale fading gain, $\beta_m(t)$ the large-scale path loss, $N_0$ the noise spectral density, and $I_m(t)$ the interference term.
The transmission and computation queues evolve as
\begin{align}
&Q_m^{\text{tx}}(t+1) = \max\!\left[ Q_m^{\text{tx}}(t) - b_m(t)T_s,\, 0 \right] + \delta_m(t) N_m(t), \\
&Q_m^{\text{comp}}(t+1) = \max\!\left[ Q_m^{\text{comp}}(t) - C_m(t)T_s,\, 0 \right] + Q_m^{\text{tx}}(t+1),
\end{align}
where $T_s$ is the scheduling interval. The corresponding average latencies are approximated via Little’s law as
\begin{equation}
\tau_m^{\text{tx}}(t) \approx \frac{Q_m^{\text{tx}}(t)}{b_m(t)}, \quad 
\tau_m^{\text{comp}}(t) \approx \frac{Q_m^{\text{comp}}(t)}{C_m(t)}.
\end{equation}
These queue dynamics form the environment’s transition function $\mathcal{P}(s'\vert s,a)$ that governs the system evolution perceived by the RL agent.

\subsection{Modeling Feedback Delay in O-RAN}

In real O-RAN systems, the effect of a control decision does not appear instantaneously in telemetry feedback. When the near-RT RIC applies an action $a_t$, its measurable impact on system KPIs (e.g., throughput, latency) is observed after a stochastic delay $d_t$. Thus, the feedback at time $t$ corresponds to the system response under an earlier state-action pair at an effective time $\ell = \max(1, t - d_t)$, where \( d_t \in \mathcal{D} = \{0, 1, \dots, D\} \) represents the delay at time \( t \) and has an unknown probability mass function \( \Pr(d_t = k) = p_k \), such that \( \sum_{k=0}^D p_k = 1 \).

\begin{figure}[t]
\centering
\includegraphics[width=3.5in]{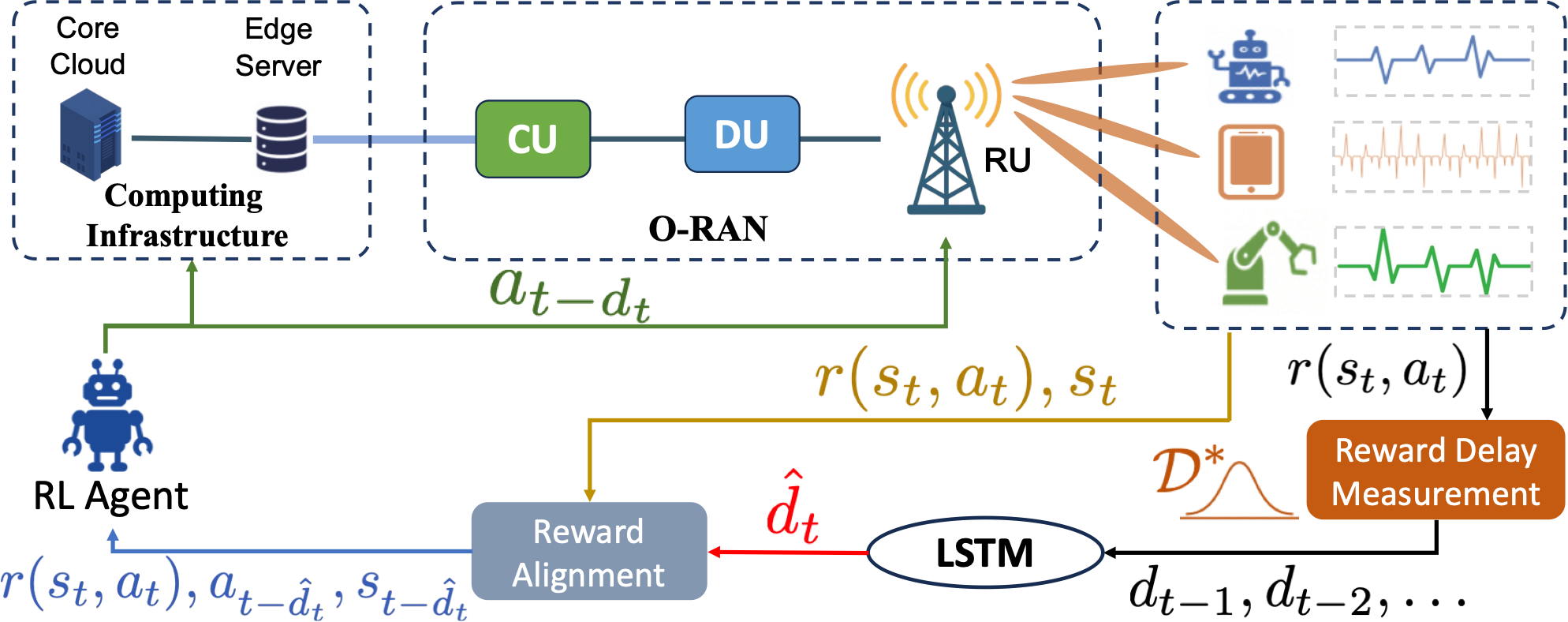}
\caption{\small Control loop in O-RAN showing stochastic feedback delay between RIC-issued actions and observed rewards across the E2 interface.}
\label{fig:architecture}
\end{figure}

This asynchronous control–feedback loop introduces a stochastic temporal misalignment between the agent’s action and its corresponding reward signal, thereby violating the standard Markov property. If unaccounted for, such delay contaminates the replay buffer and destabilizes policy convergence.
To enable delay-aware learning, the system must explicitly model this stochastic feedback process within the MDP formulation. The overall control and feedback process in the O-RAN architecture, showing the stochastic reward delay experienced across the E2 interface between the near-RT RIC and the O-DU/O-RU, is illustrated in Fig. \ref{fig:architecture}.


\subsection{Resource Update and Reliability Definition}

For each resource type $R_m(t)\in\{B_m(t),C_m(t)\}$, the near-RT RIC applies incremental updates of the form 
\begin{equation}
R_m(t)=R_m(t-1)+\rho R\phi^t\frac{\hat{R}_m(t)}{\sum_m \hat{R}_m(t)},
\end{equation}
where $\rho$ is a step-size coefficient, $\phi^t$ is the action increment chosen by the policy, and $\hat{R}_m(t)$ denotes the normalized demand estimate computed as:
\begin{align}
\hat{C}_m(t)&=\frac{\delta_m(l)N_m(l)+(\tau_m^{\text{comp}}(l-1)-T_s)^+C_m(l-1)}{\tau_m^{\max}},\\
\hat{B}_m(t)&=\frac{\delta_m(l)N_m(l)+(\tau_m^\text{tx}(l-1)-T_s)^+b_m(l-1)}{\tau_m^{\max}},
\end{align}
with $(x)^+ = \max(0,x)$. A minimum-adjustment mechanism rescales the allocations to satisfy the overall capacity constraints.
The reliability for UE $m$ is defined as the probability that latency remains below its threshold:
\begin{equation}
\mathbb{P}(\tau_m(t)\le\tau_m^{\max}) = \sum_{k} p_k \Pr[\tau_m(t)\le\tau_m^{\max}\mid d_t=k].
\end{equation}
The system aims to maintain $\mathbb{P}(\tau_m(t)\le\tau_m^{\max})\ge\varrho$ for all $m\in\mathcal{M}$.

\subsection{Reward Function and Optimization Objective}

The instantaneous reward reflects both latency satisfaction and efficient resource utilization:
\begin{align}
r(s_t,a_t)=\frac{1}{M}\sum_{m\in\mathcal{M}}\Big[
2F_a^b(\tau_m^{\max}-\tau_m(t))
+\nu F_c^d\big(\sum_m r_m(t)\big)\Big],
\end{align}
where $F_a^b(\cdot) : \mathbb{R} \rightarrow [0,1]$ and $F_c^d(\cdot) : \mathbb{R} \rightarrow [0,1]$ are smooth, strictly increasing functions parameterized by $(a,b)$ and $(c,d)$ respectively to approximate the indicator function, and $\nu \in [-1,0]$ penalizes over-provisioning. $r_m(t)$ is the normalized resource utilization, evaluated as 
$r_m(t) = \frac{B_m(t)}{2B} + \frac{C_m(t)}{2C}$.
The agent seeks to learn a stationary policy $\pi(a_t \vert s_t)$ that maximizes the expected discounted cumulative reward under stochastic feedback delays.

\section{Delay-Aware Deep Reinforcement Learning Framework}

Building on the stochastic delayed-reward formulation in Section~II, this section presents the proposed \emph{Delay-Reward Alignment} (DRA) framework for DRL in O-RAN control. The goal is to ensure robust and temporally consistent policy learning when the reward corresponding to each control action is received after a stochastic delay across the E2 interface.

\subsection{MDP Formulation with Unknown Delayed Feedback}

The environment is modeled as a Markov decision process with stochastic delayed feedback using the tuple $(\mathcal{S}, \mathcal{A}, \mathcal{P}, \mathcal{R}_d, \gamma)$, where $\mathcal{S}$ and $\mathcal{A}$ denote the state and action spaces, $\mathcal{P}(s'|s,a)$ the transition kernel, $\mathcal{R}_d$ the delay-dependent reward function, and $\gamma\!\in\![0,1)$ the discount factor.  
The stochastic delay horizon $k=d_t$ is drawn from an unknown latent distribution $\mathcal{D}^*$, and the expected cumulative delayed reward is
\begin{equation}
\label{reward}
\mathcal{R}_d(\tau_{t-k:t})
=\mathbb{E}_{k\sim\mathcal{D}^*}\!\left[\sum_{i=0}^{k}\gamma^{i}r(s_{t+i-k},a_{t+i-k})\right].
\end{equation}
This formulation generalizes standard off-policy learning by embedding unknown stochastic delay in the reward generation process.

\vspace{0.25em}
\subsection{State, Action, and Reward Specification}

At each decision epoch $t$, the near-RT RIC observes 
$
s_t = \big[ C, B, \tfrac{1}{M}\textstyle\sum_{m}C_m(t-1), 
\tfrac{1}{M}\textstyle\sum_{m}B_m(t-1),
\tfrac{1}{M}\textstyle\sum_{m}\!\text{SINR}_m^{t-1}, \tfrac{1}{M}\textstyle\sum_{m}\!N_m(t-1)\delta_m(t-1), \\ \tfrac{1}{M}\textstyle\sum_{m}\!\tau_m(t-1)\big],
$
where $\text{SINR}_m^{t} = \tfrac{P \beta_m(t)|h_m(t)|^2}{B_m(t)N_0 + I_m(t)}$.  
The agent’s action $a_t$ updates the bandwidth and computational allocations $\{B_m(t),C_m(t)\}$ for all $m\!\in\!\mathcal{M}$, subject to total resource constraints $\sum_m B_m(t)\!\leq\!B$ and $\sum_m C_m(t)\!\leq\!C$.
The instantaneous reward follows the definition in~\eqref{reward}, combining latency satisfaction and utilization efficiency, while the realized delayed reward at time $t$ is as follows:
\(
\mathcal{R}_d^t(\tau_{t-k:t}\mid k)
= \sum_{i=0}^{k}\gamma^i r(s_{t+i-k},a_{t+i-k}).
\)

\vspace{0.25em}
\subsection{Delay Prediction and Reward Realignment}
\begin{figure}[t!]
    \centering
    \vspace{-0.in}
    \includegraphics[width = 3.55in]{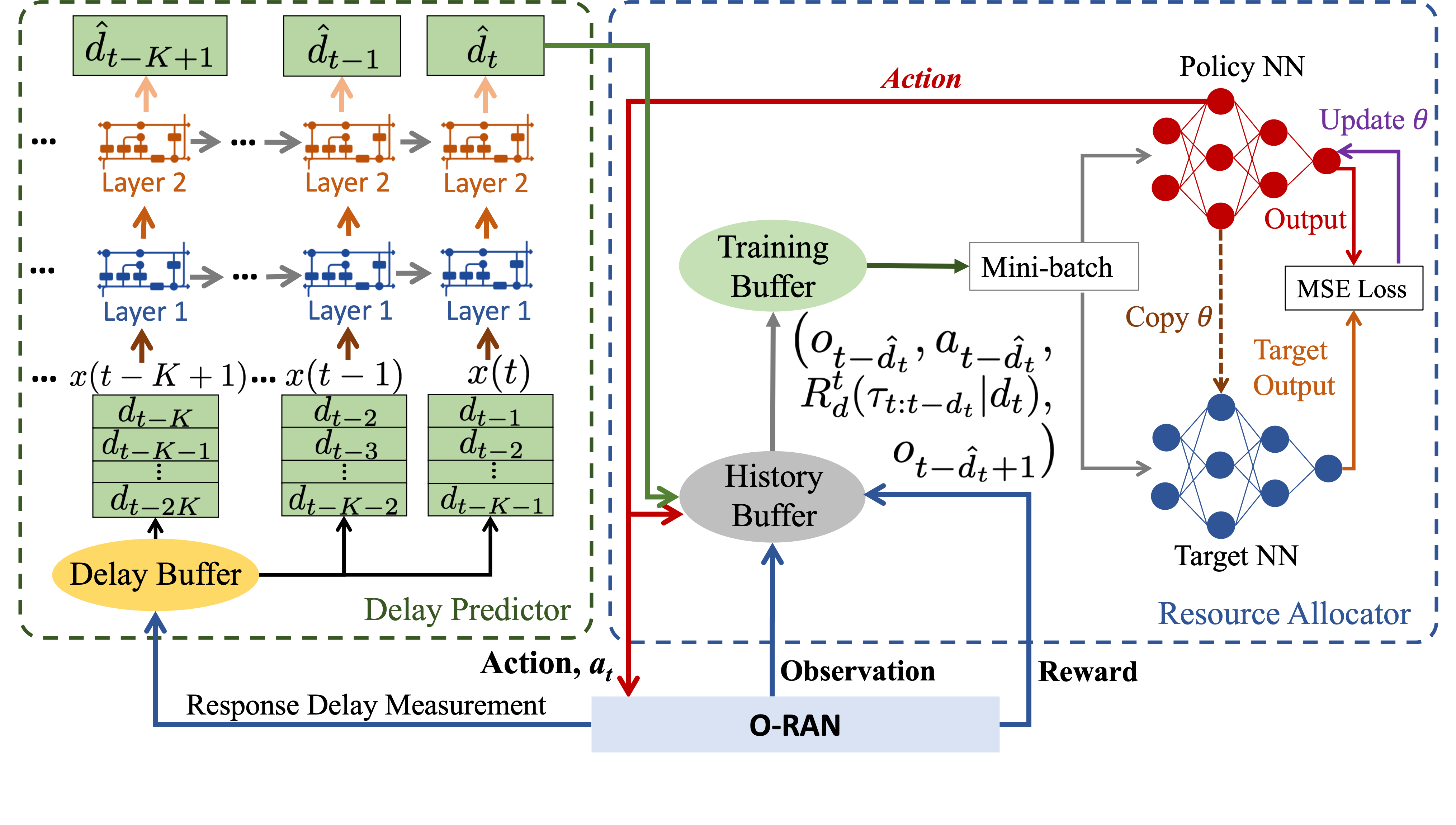}
    \caption{\small Integrated training architecture for mitigating delayed reward corruption using LSTM-driven delay prediction and off-policy DRL. \vspace{-0in}}
    \label{fig:lstm}
\end{figure}

Fig. \ref{fig:lstm} illustrates the integrated DRA training architecture that combines an LSTM-based delay predictor with an off-policy DRL orchestrator.
To mitigate policy corruption caused by delayed rewards, the DRA framework integrates a DRL-based orchestrator and an LSTM-based delay predictor operating under strict modular separation. Three buffers are maintained:  
(1) $\mathcal{H}$ for raw trajectories,  
(2) $\mathcal{G}$ for observed delay history $\{k_t\}$, and  
(3) $\mathcal{F}$ for corrected replay transitions constructed using the predicted delay $\hat{k}_t$.

At each training iteration, the predictor uses past delay samples $\mathcal{K}=\{k_{t-1},\dots,k_{t-K}\}$ to estimate $\hat{k}_t$, reassigning the delayed reward $\mathcal{R}_d^t$ to the appropriate historical $(s_{t-\hat{k}_t},a_{t-\hat{k}_t})$ pair.  
This temporally aligned transition is inserted into $\mathcal{F}$, ensuring high-purity experiences for policy learning.
The LSTM predictor is trained to minimize the mean-squared delay prediction error,
\begin{equation}
\mathcal{L}_{\text{delay}}=\tfrac{1}{|\mathcal{M}|}\!\sum_{t\in\mathcal{M}}\!(d_t-\hat{d}_t)^2,
\end{equation}
while the DRL agent minimizes the mean-squared Bellman error:
\begin{equation}
L(\theta)=\mathbb{E}\big[(Q_\theta(s_{t-k},a_{t-k})-y_t)^2\big],
\end{equation}
where $y_t=\hat{r}_t+\gamma Q_{\theta^-}(s_{t-k+1},\pi_{\theta^-}(s_{t-k+1}))$ and $\hat{r}_t=r_{t+\hat{k}_t}$.

\subsection{Online Training Procedure}

The full online training process is summarized in Algorithm~\ref{alg:lstm_drl}, which alternates between environment interaction, delay prediction, and policy update phases. This design decouples delay estimation from reinforcement updates, thereby preserving gradient stability under stochastic and non-Markovian delays. The resulting formulation is compatible with standard off-policy algorithms such as DQN, DDPG, and SAC, providing a plug-and-play extension for delay-aware training in near-RT RIC implementations.

\begin{algorithm}[t]
\small
\caption{\small LSTM-Enhanced Off-Policy DRL for Delay-Aware Resource Orchestration}
\label{alg:lstm_drl}
\begin{algorithmic}[1]
\State \textbf{Initialize:} environment $\mathcal{P}_\sigma$, replay buffers $\mathcal{H}$, $\mathcal{F}$, $\mathcal{G}$; policy $\pi_\theta$, critic $Q_\phi$, and LSTM predictor $\mathcal{F}_\psi$
\For{each episode}
  \For{each time step $t$}
    \State Observe $s_t$; choose $a_t\!\sim\!\pi_\theta(s_t)$; apply $a_t$ to update $\{B_m,C_m\}$ 
    \State Observe delayed reward $\mathcal{R}_d^t(\tau_{t-k_t:t}\mid k_t)$, where $k_t\!\sim\!\mathcal{D}$
    \State Update delay predictor using past $\{k_i\}$ and estimate $\hat{k}_t$
    \State Construct corrected tuple $(s_{t-\hat{k}_t},a_{t-\hat{k}_t},\mathcal{R}_d^t,s_{t-\hat{k}_t+1})$
    \State Store aligned tuple in $\mathcal{F}$ and true delay $k_t$ in $\mathcal{G}$
    \State Sample minibatch from $\mathcal{F}$; update $Q_\phi$ and $\pi_\theta$ via gradient descent
  \EndFor
\EndFor
\State \textbf{Return:} optimized delay-aware policy $\pi^*$
\end{algorithmic}
\end{algorithm}

\section{Experimental Evaluation}

This section evaluates the proposed DRA mechanism integrated into three popular off-policy RL algorithms, i.e., Deep Q-Network (DQN), Deep Deterministic Policy Gradient (DDPG), and Soft Actor–Critic (SAC), for O-RAN control under stochastic reward delay. The experiments employ empirical delay traces from the OAIC platform, enabling realistic modeling of control–feedback latency between the near-RT RIC and O-DU.

\subsection{Experimental Setup}

Our experimental setup leveraged the OAIC testbed~\cite{oaic} and Python 3.7 with PyTorch 1.10.1 running on an NVIDIA GeForce MX350 GPU (8GB VRAM). 
Delay-response measurements were recorded from real-time OAIC experiments and discretized into time slots for training and evaluation. A delay 
model of the following form was empirically fitted on the empirical measurements:
$d_t=\min \{D, X_t + A \sin(2\pi f t ) + \Phi \}$,
where $X_t \sim \text{Exp}(1/\lambda)$ with mean $1 /\lambda$ seconds , while $A$, $f$, and $\Phi$ specify the amplitude, frequency, and constant shift of the periodic component, respectively. 
The total training cycle, including baseline and proposed models, spanned approximately 48 hours. Table~\ref{tab:SOTA} summarizes the key environment and learning parameters. For reward shaping, we employed sigmoid functions of the form \( F_a^b(x) = \frac{1}{1 + \exp(-a(x - b))} \), where parameters \( a \) and \( b \) tune the sharpness and shift of the response. This flexibility enabled precise alignment of the reward signal with critical latency and utilization thresholds.

\begin{table}[t]
\centering
\footnotesize
\renewcommand{\arraystretch}{1.1} 
\begin{tabularx}{0.47\textwidth}
{
| >{\centering\arraybackslash}X
| >{\centering\arraybackslash}X
| >{\centering\arraybackslash}X
| >{\centering\arraybackslash}X |
}
\hline \textbf{Lr Params} & \textbf{Value} & \textbf{Env Params} & \textbf{Value} \\
\hline $K$  & 15 & $M$, $d_m^{(t)}$ (km) & 4$\sim$9, [0.4, 0.5]\\
\hline $\alpha_{\text{soft-update}}$& 1e-3 & $C$ (Mbps) & $1000$ \\
\hline $N_s^{1}, N_s^{2}, N_s^{3}$ & 10, 10, 10 & $f_c$ (GHz) & 3 \\
\hline$\mu_{\text{DRL}}$, $\mu_{\text{LSTM}}$ & 5e-5, 1e-3 & $\delta_m(t)$ (Mb) & [0.003, 0.3] \\
\hline$S_{\text{batch}}^{\text{DRL}}$ & 256 & $\psi_{\xi}$ (dB) & 3.5 \\
\hline$S_{\text{batch}}^{\text{LSTM}}$& 32 & $T_n$,  $T_s$ (s)& 150, 0.2 \\
\hline$H_{\text{DDPG}}$ & $\{64, 64\}$ & $\tau_{\text{e}}$ (s), $\tau_{\text{u}}$ (s) & 0.1, 0.001 \\
\hline $H_{\text{DQN}}$ & $\{25, 25\}$ & $P$ (dBm) & 32.3 \\
\hline $H_{\text{SAC}}$ & $\{64, 64\}$ &$B$ (MHz) & $1000$ \\
\hline
$a$, $b$ & 20, 5e-4& $N_m(t)$ & [30, 55]  \\
\hline $\nu$,  $\varrho$, $\gamma$, & -0.15, 0.99, 0.1 & $\eta$ (dB) & 0.1\\
\hline $H_{\text{LSTM}}$ & $\{64, 64\}$ & $N_0$ (dBm/Hz) & -163 \\
\hline
$S_{\text{buffer}}$& 5e5& $f$, $\Phi$ & 0.1/2$\pi$, [3, 8]\\
\hline
$c$, $d$& 10, 0.6 & $\lambda$, $A$& [0.4, 1], [0, 5]\\
\hline
\end{tabularx}
\caption{\label{tab:SOTA}Simulation Parameters.\vspace{-0in}}
\end{table}

\begin{figure}[t]
\vspace{-0in}
    \centering
    \begin{subfigure}[b]{0.235\textwidth}
        \includegraphics[width=\textwidth]{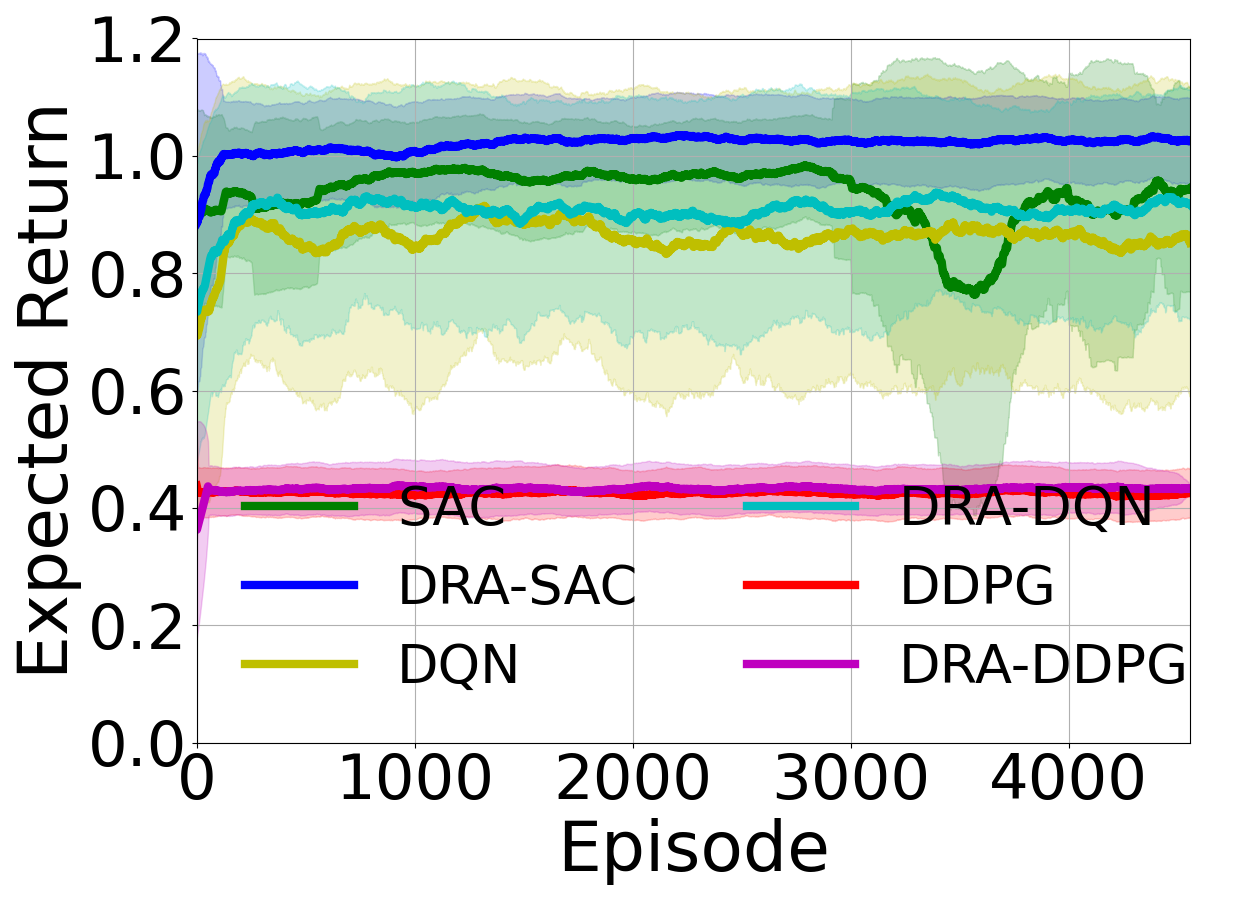}
        \caption{Expected Return}
    \end{subfigure}
    \hfill
    \begin{subfigure}[b]{0.235\textwidth}
        \includegraphics[width=\textwidth]{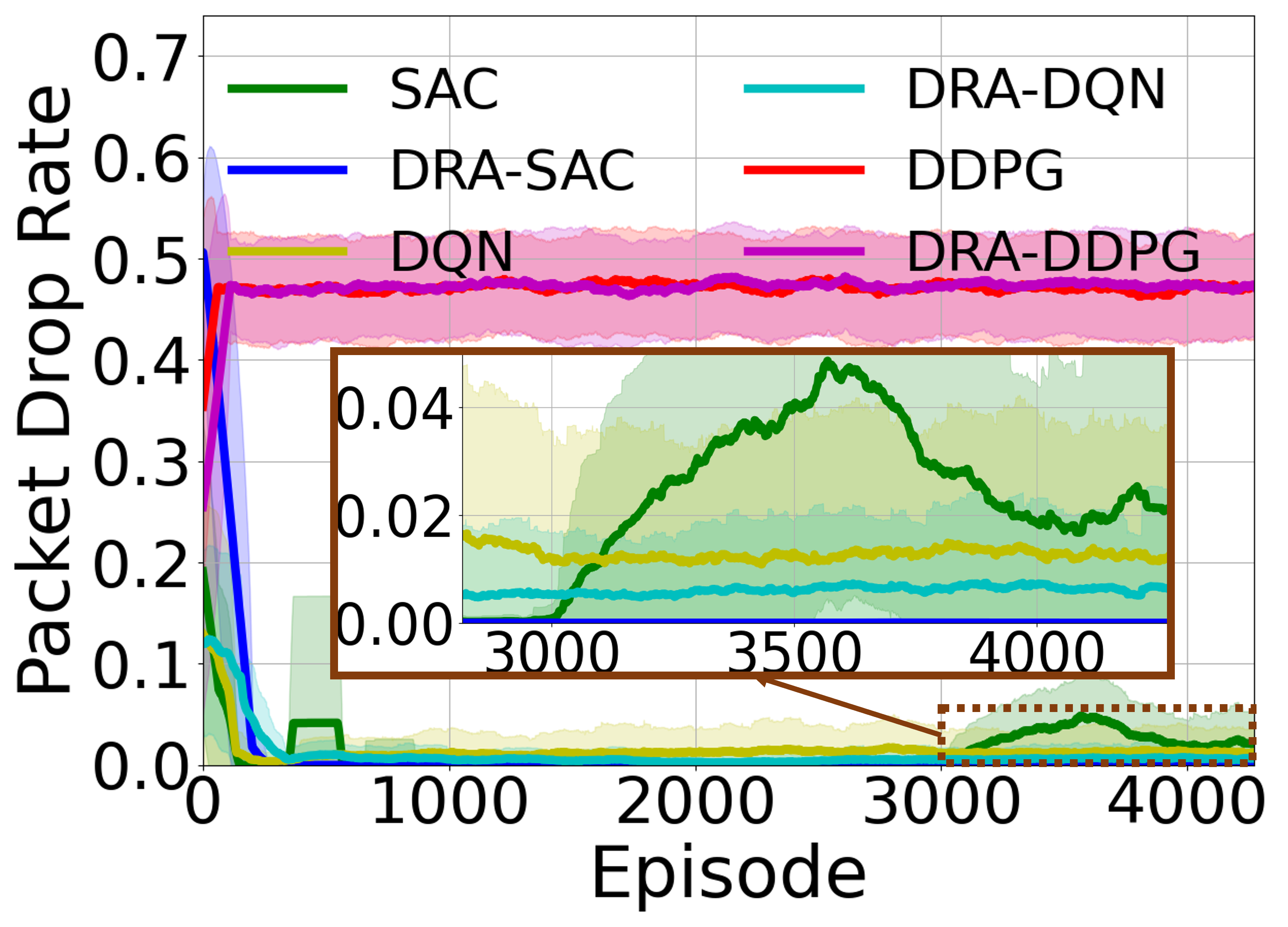}
        \caption{Packet Drop Ratio}
    \end{subfigure}
    \caption{Training rewards of resource orchestration agents under varying strategies.\vspace{-0.0in}}
    \label{fig:training_reward}
\end{figure}

\begin{figure}[t!]
    \centering
    \begin{subfigure}[b]{0.235\textwidth}
        \includegraphics[width=\textwidth]{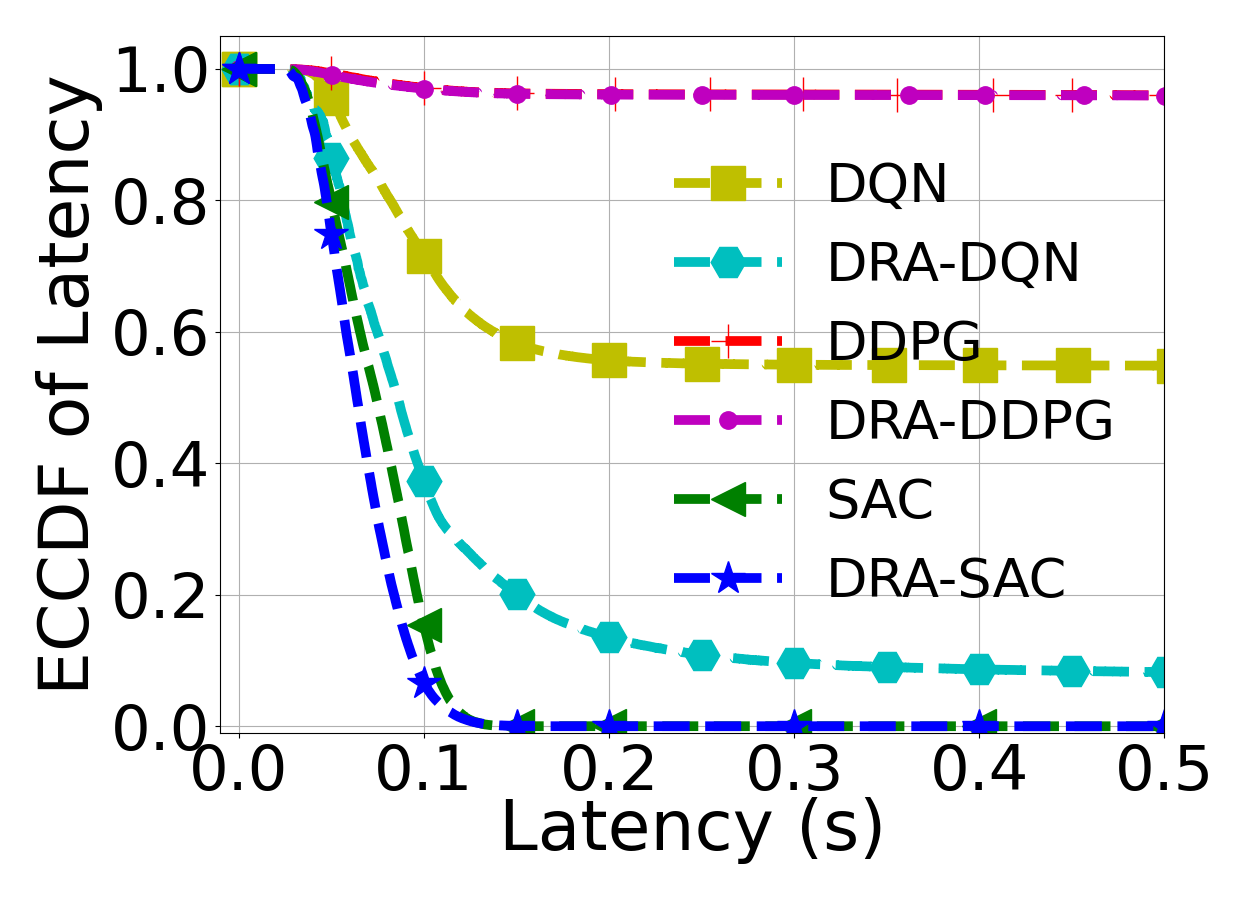}
        \caption{eMBB Slice}
    \end{subfigure}
    \hfill
    \begin{subfigure}[b]{0.235\textwidth}
        \includegraphics[width=\textwidth]{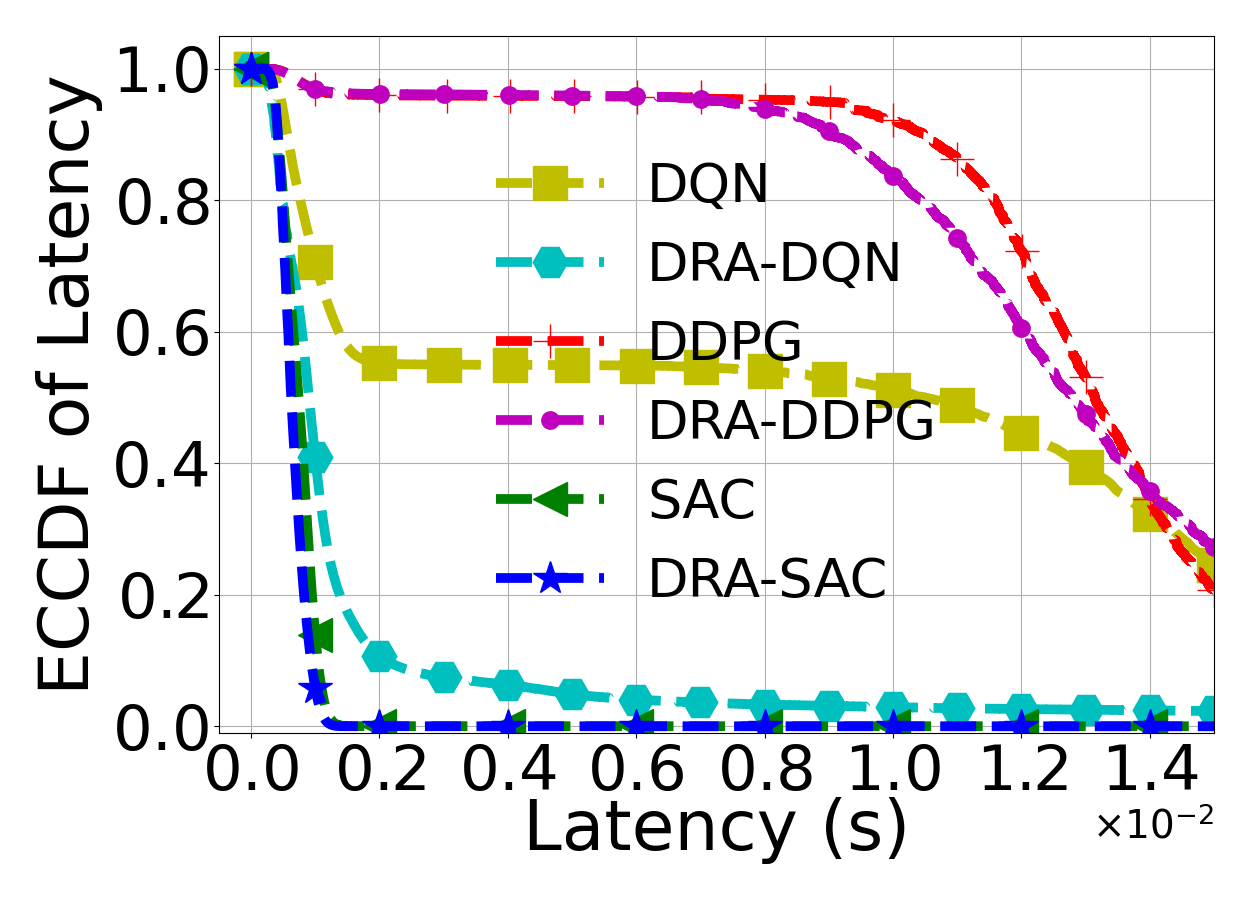}
        \caption{URLLC Slice}
    \end{subfigure}
    \caption{\small ECCDF of per-slice latency for DQN, SAC, and DDPG with and without DRA compensation.
    \vspace{-0.0in}}
    \label{fig:eccdfs}
\end{figure}

For each algorithm (DQN, SAC, and DDPG), we train two variants:
\begin{itemize}
    \item \textbf{Baseline:} the standard algorithm without delay compensation;
    \item \textbf{DRA-variant:} the same algorithm enhanced with the proposed DRA framework described in Section~III.
\end{itemize}
All agents are trained for $5\times10^4$ steps with discount factor $\gamma = 0.95$, learning rate $10^{-4}$, and batch size 64. The LSTM-based delay predictor within DRA uses a single hidden layer with 64 units and a temporal window of 10 steps.

Performance is evaluated using the following metrics relevant to near-RT RIC operation: (i) \textit{Average episodic return}, indicating policy convergence stability; (ii) \textit{Latency violation ratio}, percentage of users exceeding $\tau_m^{\max}$; (iii) \textit{Packet drop ratio}, fraction of packets not delivered within deadlines; and (iv) \textit{Resource efficiency}, throughput-to-bandwidth utilization ratio.

\begin{figure}[t]
    \centering
    \begin{subfigure}[b]{0.235\textwidth}
        \includegraphics[width=\textwidth]{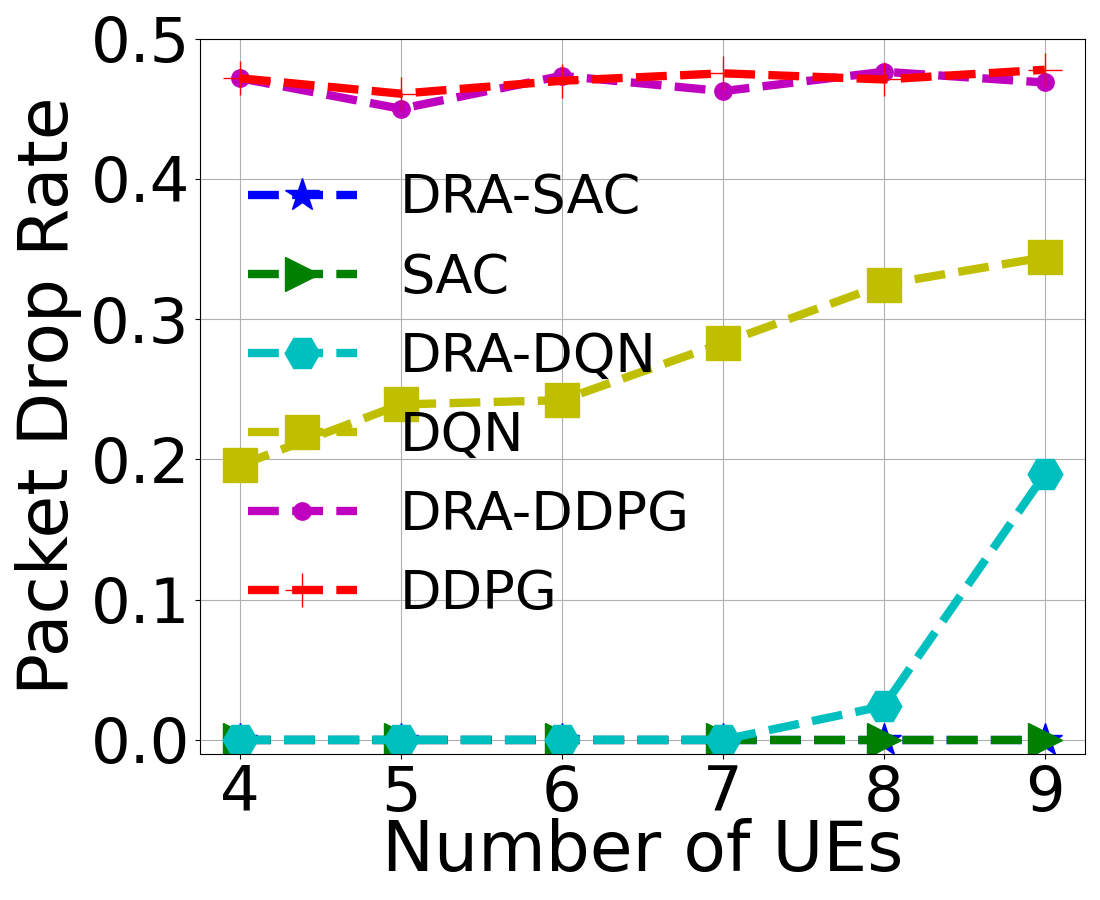}
        \caption{Packet Drop Ratio}
    \end{subfigure}
    \hfill
    \begin{subfigure}[b]{0.235\textwidth}
        \includegraphics[width=\textwidth]{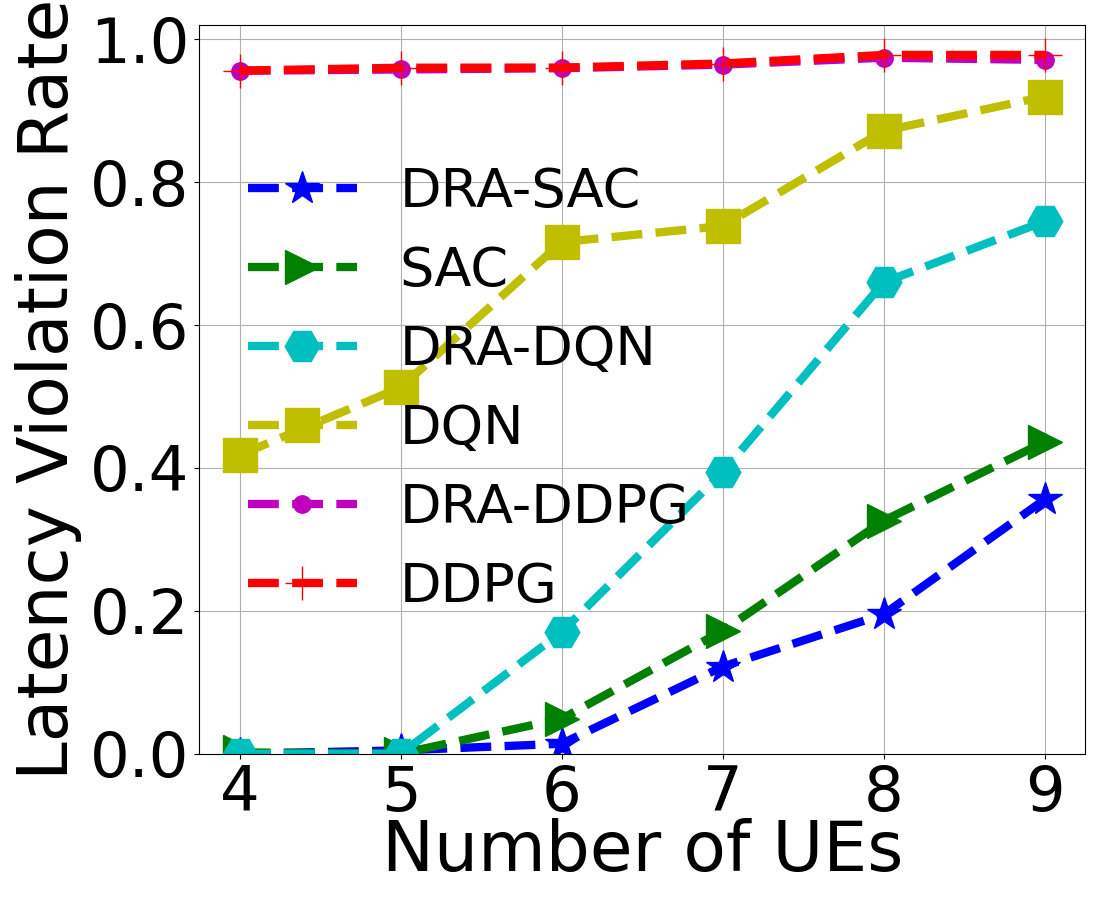}
        \caption{Latency Violation Rate}
    \end{subfigure}
    \label{fig:util-across-pro}
    \begin{subfigure}[b]{0.235\textwidth}
        \includegraphics[width=\textwidth]{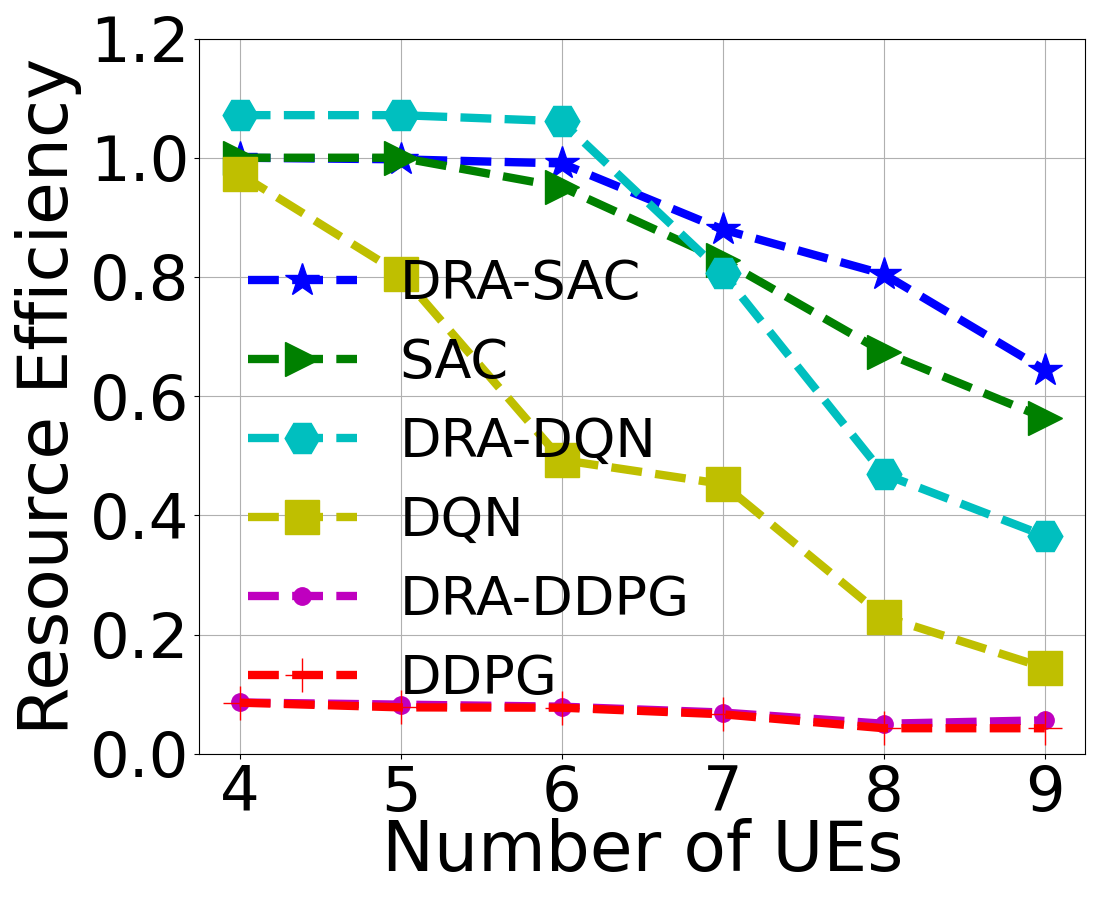}
        \caption{Resource Efficiency}
    \end{subfigure}
    \hfill
    \begin{subfigure}[b]{0.235\textwidth}
        \includegraphics[width=\textwidth]{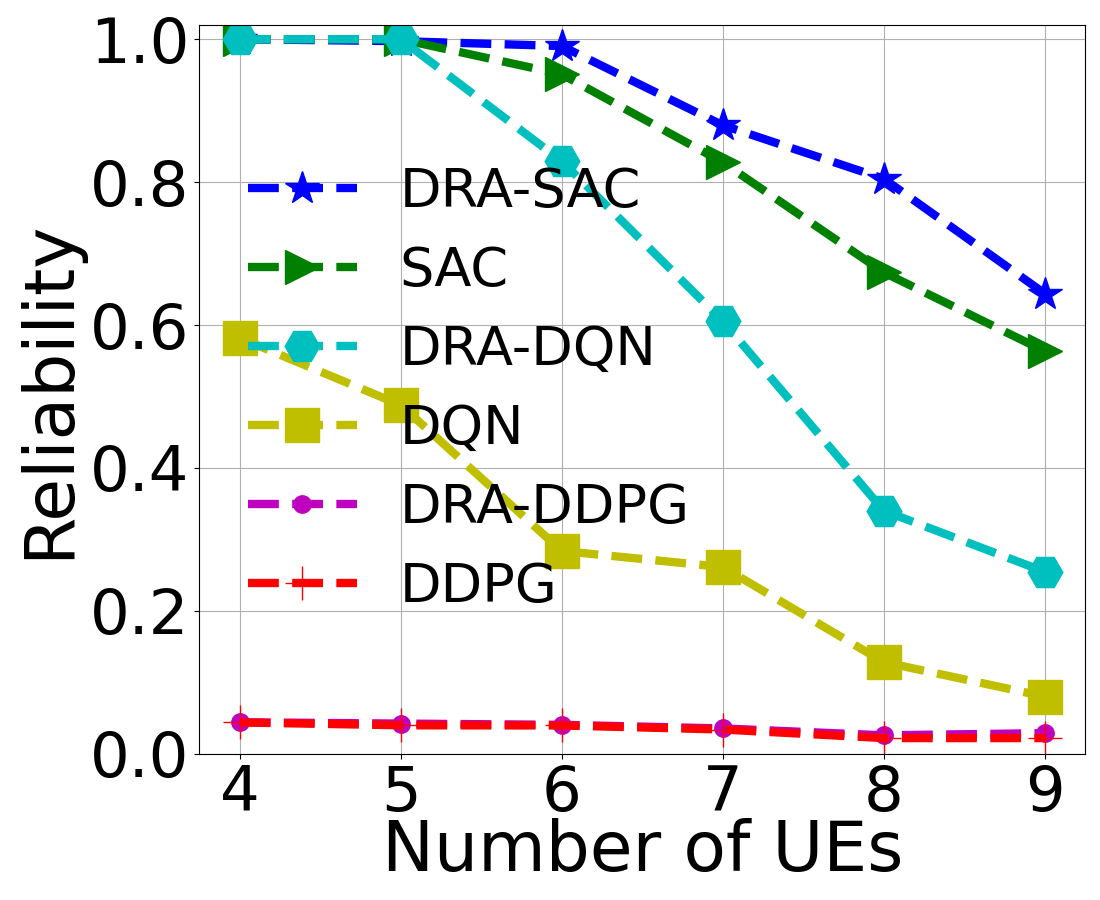}
        \caption{Reliability}
    \end{subfigure}
    \caption{\small Aggregate QoS metrics for increasing UE count comparing DQN, SAC, DDPG, and their DRA-enhanced counterparts.\vspace{-0in}}
    \label{fig:test_KPIs}
\end{figure}

\subsection{Results and Discussion}

Fig.~\ref{fig:training_reward} compares the convergence behavior of all six agents. In Fig.~\ref{fig:training_reward}(a), the DRA-enhanced algorithms (DRA-DQN, DRA-SAC, DRA-DDPG) exhibit smooth and monotonic improvement in expected return, whereas their non-DRA counterparts show slower or unstable convergence when stochastic delay is present. The advantage of the DRA mechanism is further evident in Fig.~\ref{fig:training_reward}(b), where all DRA variants maintain substantially lower packet-drop ratios across episodes, highlighting improved temporal credit assignment under delayed feedback.

The end-to-end latency distributions are characterized by the empirical complementary cumulative distribution functions (ECCDFs) shown in Fig.~\ref{fig:eccdfs}. For eMBB users, as shown in Fig.~\ref{fig:eccdfs}(a), DRA-augmented policies achieve tighter latency concentration, with 95th-percentile delays below 15~ms. For URLLC users, as shown in Fig.~\ref{fig:eccdfs}(b), only the DRA variants maintain sub-1~ms latency at reliability above $99.9\%$, while standard baselines exhibit heavy tails due to uncorrected reward misalignment. These results confirm that DRA preserves QoS isolation across heterogeneous slices even under feedback uncertainty.

The system-level metrics in Fig.~\ref{fig:test_KPIs} further reinforce this observation. Across increasing UE density, the DRA variants consistently achieve lower latency-violation and packet-drop ratios, as shown in Figs.~\ref{fig:test_KPIs}(a)–(b), as well as higher resource efficiency and reliability shown in Figs.~\ref{fig:test_KPIs}(c)–(d). Among all, DRA-SAC delivers the best trade-off between learning stability and resource efficiency, followed by DRA-DDPG and DRA-DQN. Quantitatively, the DRA framework reduces latency violations by up to 25\% and improves spectral–computational efficiency by 12\% compared with the non-compensated baselines.

\section{Conclusion}

This letter addressed the challenge of stochastic reward delay in reinforcement learning (RL)-based control for O-RAN systems. We formulated the O-RAN resource allocation problem as an MDP with delayed rewards and proposed a delay-aware deep RL framework that integrates an LSTM-based delay predictor for temporal reward realignment. Using empirical delay traces derived from the OAIC testbed, the results show that the proposed method achieves faster convergence, higher reliability, and better resource efficiency than conventional RL baselines. Future work will explore multi-agent extensions for cooperative delay compensation and on-device deployment within near-RT RIC platforms.

\vspace{-0.0in}
\bibliographystyle{IEEEtran}
\bibliography{IEEEabrv,Bibliography}

\end{document}